\documentclass[aps,prb,twocolumn,superscriptaddress,citeautoscript]{revtex4}
\usepackage{graphicx,textcomp}
\usepackage[version=3]{mhchem}
\begin{document}
\title{Pseudo-isotropic upper critical field in cobalt-doped SrFe$_2$As$_2$ epitaxial films}
\author{S.~A.~Baily}
\email[]{sbaily@lanl.gov}
\affiliation{MPA-NHMFL, Los Alamos National Laboratory, Los Alamos, NM, 87545, USA}
\affiliation{MPA-STC, Los Alamos National Laboratory, Los Alamos, NM, 87545, USA}
\author{Y.~Kohama}
\affiliation{MPA-NHMFL, Los Alamos National Laboratory, Los Alamos, NM, 87545, USA}
\affiliation{Materials and Structures Laboratory, Tokyo Institute of Technology, Yokohama, 226-8503, Japan}
\author{H.~Hiramatsu}
\affiliation{ERATO-SORST, Japan Science and Technology Agency in Frontier Research Center, Tokyo Institute of Technology, Yokohama, 226-8503, Japan}
\author{B.~Maiorov}
\affiliation{MPA-STC, Los Alamos National Laboratory, Los Alamos, NM, 87545, USA}
\author{F.~F.~Balakirev}
\affiliation{MPA-NHMFL, Los Alamos National Laboratory, Los Alamos, NM, 87545, USA}
\author{M. Hirano}
\author{H. Hosono}
\affiliation{ERATO-SORST, Japan Science and Technology Agency in Frontier Research Center, Tokyo Institute of Technology, Yokohama, 226-8503, Japan}
\affiliation{Frontier Research Center, Tokyo Institute of Technology, Yokohama, 226-8503, Japan}
\date{\today}
\begin{abstract}
The temperature and angular dependence of the upper critical field ($H_{c2}$) is reported for cobalt-doped SrFe$_2$As$_2$ epitaxial films between $T_c$ and 0.5~K in pulsed magnetic fields up to 50~T. For $H\parallel c$, $H_{c2}$ is close to a linear function of temperature, while in the perpendicular direction there is significant downward curvature that results in an $H_{c2}$ ratio ($\gamma=H_{c2}^{\perp}/H_{c2}^{\parallel}$) that decreases nearly linearly with temperature, approaching $\gamma=1$ at low temperature with $H_{c2}(0)=47$~T. We measure the complete $H_{c2}$ phase diagram including angular dependence and model the data using a two band theory allowing us to determine the anisotropy of both bands, their relative diffusivities, and the relationship between BCS coupling constant matrix elements. We find an unusual relationship between the diffusivities of the two bands, with two anisotropic and opposite bands. This relationship is supported by the observation of a local maximum for $H_{c2}^{\parallel}$ at low temperature.
%We present resistivity measurements of the complete superconducting upper critical field ($H_{c2}$) phase diagram as a function of angle ($\theta$) and temperature ($T$) for cobalt-doped SrFe$_2$As$_2$ epitaxial films to 0.5 K and 50 T. Although $H_{c2}(\theta)$ at 10~K is indistinguishable from that derived from a single-band anisotropy model, the apparent anisotropy $H_{c2}^{\perp c}/H_{c2}^{\parallel c}$ linearly decreases to 1 at low T, with $H_{c2}(0)=47$~T. The data is well described by a two-band model with small, opposing anisotropies for the bands. This unusual relationship is confirmed by the observation of a local maximum for $H_{c2}^{\parallel c}$ at low $T$.
\end{abstract}
\maketitle

The discovery of a new class of superconductors \cite{YKamihara2008} has regenerated interest in high temperature superconductivity. Iron arsenides become superconducting when doped with either holes or electrons, and multiple sites are available for doping \cite{MRotter2008,GCao2008}. This opportunity to tune these materials in many ways without destroying superconductivity is a tremendous advantage because specific aspects of theories can be tested, potentially allowing one to reveal the mechanism of high temperature superconductivity. The relatively high upper critical fields ($H_{c2}$) with atypical temperature dependence in these materials immediately attracted much interest \cite{FHunte2008,YKohama2008a,YKohama2008b,JJaroszynski2008a}.

Early on, the curvature observed in the temperature dependence of $H_{c2}$ in these materials made the dirty two band model \cite{AGurevich2003,AAGolubov2003} used to describe \ce{MgB2} a natural and useful choice \cite{FHunte2008,JJaroszynski2008a}. Already, investigations of single crystals have revealed an unusual temperature-dependent anisotropy and confirmed that a multi-band description was necessary \cite{HQYuan2008,JJaroszynski2008b}. Now our ability to measure epitaxial films enables the study of the effects of strain, defects, and dimensionality as well as facilitating high precision electrical transport measurements. While many properties of \ce{YBa2Cu3O7} superconductors can be scaled with angle using a single band temperature-independent electronic mass anisotropy, the two-band model predicts a richer angular dependence that may be more sharply peaked or even have two maxima \cite{AGurevich2003}, with the shape and degree of anisotropy changing with temperature and the details of both bands. Therefore, just as angular dependence measurements provided great insight to the nature of vortex states \cite{SABaily2008}, they are also key experiments to capture the richness predicted from the competition between bands in multiband superconductors. Since the iron arsenides are not as well studied as \ce{MgB2}, many assumptions must be made about the band structure \cite{VBZabolotnyy2008,CLiu2008,DJSingh2008}. By measuring $H_{c2}$ over the complete temperature and angular range one can test these assumptions and narrow the range of feasible parameters, leaving few ambiguities. We include measurements of the angular dependence of $H_{c2}$ at low temperature and find this regime is crucial to understanding the behavior of these materials.

Most research on $H_{c2}$ behavior has focused on LnFeAsO (lanthanide) compounds doped at the oxygen site \cite{JJaroszynski2008b,YJia2008,SWeyeneth2008}, or hole-doped \ce{AEFe2As2} (AE=Alkali Earth) compounds \cite{HQYuan2008,HJKim2008}. By achieving electron doping in Co-doped \ce{SrFe2As2}, we hope to learn more about the less well studied band. In this letter, we report the temperature and angle dependence of the resistive $H_{c2}$ for Co-doped \ce{SrFe2As2} epitaxial films, for which the upper critical fields at low temperature are accessible to pulsed field magnets. Although significant quantitative differences exist, this composition still shares physics with the electron-doped oxypnictides that have higher $H_{c2}(0)$ and $T_c$ values. The use of epitaxial thin films permits an excellent signal-to-noise ratio as well as allowing us to ascertain the effects of defects and disorder (which for high temperature superconductors may differ greatly for single crystals and films).

\begin{figure}
\includegraphics[width=8.6cm]{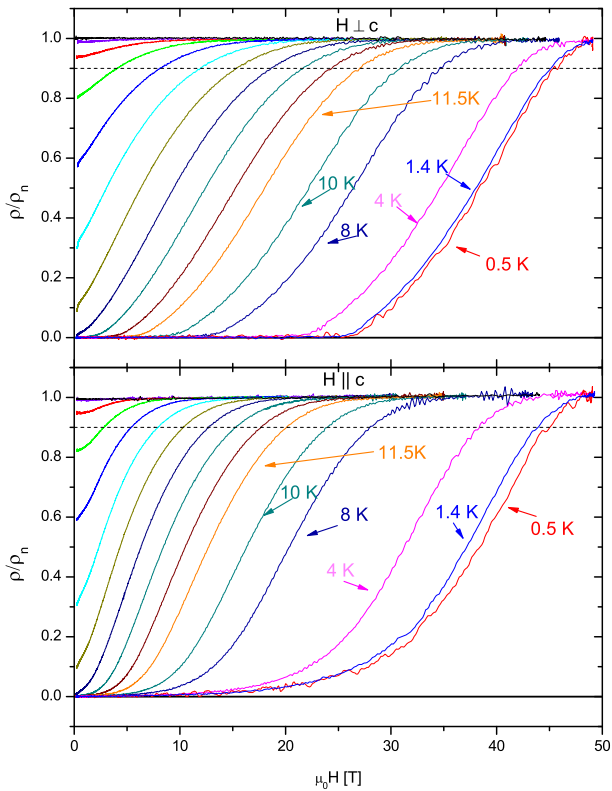}
\caption{(Color online) Normalized resistivity vs.\ applied magnetic field as a function of temperature for $H \perp c$ and $H \parallel c$. The unlabeled curves are separated by 1~K, from 11.5~K to 20.5~K. Data for 30~K up to 30~T are also superimposed.\label{RvsT}}
\end{figure}

The Co-doped \ce{SrFe2As2} epitaxial films were grown at 670\textcelsius\ on \ce{(La , Sr)(Al , Ta)O3} substrates via a pulsed laser deposition system\cite{HHiramatsu2008b} from a \ce{SrFe_{1.8}Co_{0.2}As2} polycrystalline target \cite{HHiramatsu2008a}. The FWHM of the out-of-plane rocking curve for the 002 diffraction peak is 1.6\textdegree . X-ray fluorescence spectrometry is consistent with the films having the same Fe/Co ratio as the target. The data shown in this letter are from a 180~nm film with $T_{c}=17.1$~K (midpoint) and a normal state resistivity $\rho_n$ of 330~$\mu\Omega$cm below 30~K.  The resistive transition width (2.7~K FWHM) is not as sharp as that of single crystals \cite{RProzorov2008}. However, both films measured show a single and smooth transition with resistivity dropping to zero. The samples were placed on a rotator stage for resistivity measurements, and a small 100~kHz ac current of $35~\textrm{A/cm}^2$ was applied perpendicular to the magnetic field and the axis of rotation. Ohmic behavior was verified by doubling and halving the current. A digital lock-in technique was used to record the voltage as a pulsed magnetic field of up to 50~T was applied. Some eddy current heating was observed for measurements in the 4--8~K range. At 4~K and below, the sample was submerged in liquid helium, preventing heating.  In the normal state, these films have nearly constant resistivity as a function of temperature below 30~K and negligible magnetoresistance regardless of the field direction, as shown in Fig.~\ref{RvsT}.

\begin{figure}
\includegraphics[width=8.6cm]{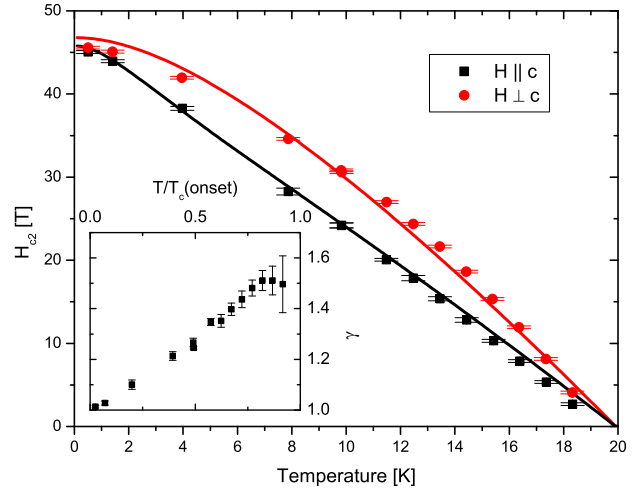}
\caption{(Color online) Upper critical field vs.\ temperature for $H \parallel c$ and $H \perp c$ as determined by a criterion of 90\% of the normal-state resistivity. Solid lines show the results of a self-consistent fit for both directions.\label{Hc2Tdep} The inset shows ($\gamma=H_{c2}^{\perp}/H_{c2}^{\parallel}$) deceasing linearly as a function of temperature.\label{anisotropy}}
\end{figure}

\begin{figure}
\includegraphics[width=8.6cm]{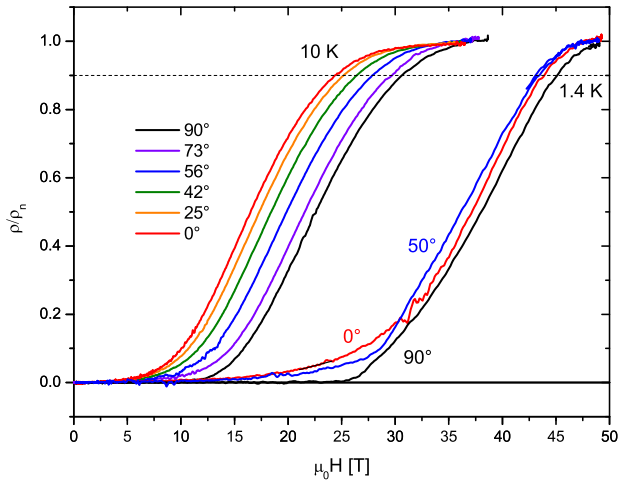}
\caption{(Color online) Normalized resistivity vs.\ applied magnetic field for angular dependence measurements at 10~K and 1.4~K.\label{angular}}
\end{figure}

The temperature dependence of $H_{c2}$ (determined by 90\% of $\rho_n$) depicted in Fig.~\ref{Hc2Tdep} has several interesting features. Estimate of errors due to variation in background signal over the angular range, determination of lock-in phase, magnetoresistance, electrical noise, possible temperature variation of $\rho_n$, and the effect of temperature variations on $H_{c2}$ are used to generate vertical errorbars. Errors in temperature and angle are negligible on the scales used in the figures. For $H \parallel c$, $H_{c2}$ is nearly linear in temperature, while for $H \perp c$, $H_{c2}$ is sharply curved with a negative second derivative with respect to temperature.  This results in an $H_{c2}$ ratio ($\gamma=H_{c2}^{\perp}/H_{c2}^{\parallel}$) which decreases with decreasing temperature and approaches 1 at 0~K (Fig.~\ref{Hc2Tdep} inset). The same qualitative behavior was observed in oxypnictide and \ce{(Ba , K)Fe2As2} single crystals \cite{HQYuan2008,JJaroszynski2008b}. This indicates that the increased disorder in thin films is not the cause of the $\gamma$ temperature behavior we report here. Indeed, a film with lower $T_c$ measured in this study showed the same behavior in $\gamma$, demonstrating that these materials are not particularly sensitive to disorder. Both the \ce{(Ba , K)Fe2As2} crystals \cite{HQYuan2008} and the films of this study show $\gamma$  near 1.5 at $0.75T_c(\textrm{onset})$, which drops to $\approx 1.3$ at $T/T_c=0.5$. However, our excellent signal-to-noise ratio in angular dependent measurements (Fig.~\ref{angular}) allows us to observe that while $\gamma$ is within 3\% of 1 below 1.4~K, Co-doped \ce{SrFe_2As_2} is not truly isotropic at low temperature. The quantitative similarities in $H_{c2}$ anisotropy of \ce{(Ba , K)Fe2As2} and \ce{Sr(Fe , Co)2As2} are remarkable because the former is hole-doped, and the latter electron-doped, and one would expect different bands to be important.

\begin{figure}
\includegraphics[width=8.6cm]{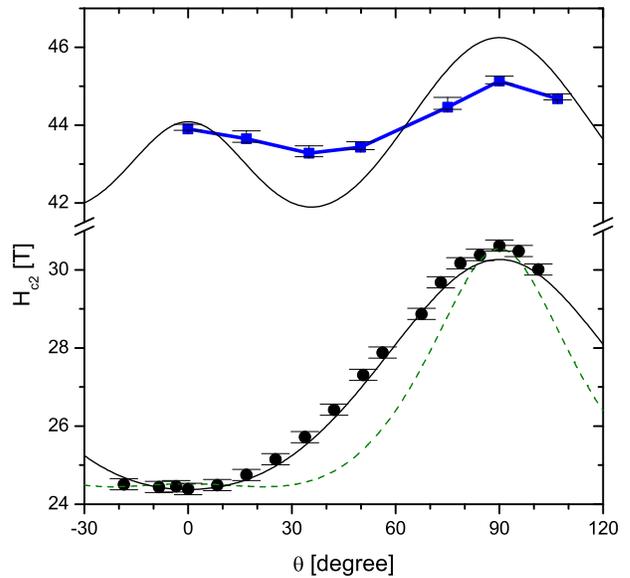}
\caption{(Color online) $H_{c2}$ angular dependence at 10~K (circles) and 1.4~K (squares) using a 90\% criterion. Solid lines show a self-consistent fit. The dashed line is a fit only constrained by the temperature dependence of $H_{c2}^{\perp}$ and $H_{c2}^{\parallel}$. \label{angularfits}}
\end{figure}

There are two ways to obtain $\gamma=1$ at low temperature within the context of a model based on band anisotropy. One is to have an isotropic band that dominates at low temperature, and the other is to have bands which have opposing anisotropies. The dirty two-band model describes the temperature and angle dependence of $H_{c2}$ in terms of the BCS coupling constants and the diffusivities of the bands\cite{AGurevich2003}. In this model, the angular dependence of the diffusivity $D(\theta)$ is described by the two coefficients for each band $m$ according to:
\begin{equation}
D_m(\theta)=\sqrt{{D_m^{(a)}}^2\cos^2(\theta)+D_m^{(a)}D_m^{(c)}\sin^2(\theta)},
\end{equation}
where $\theta$ is the angle between the applied field and the crystallographic c-axis. $H_{c2}(\theta,T)$ is given by the solution to the equation:
\begin{eqnarray}
\frac{2\lambda_{11}\lambda_{22}-2\lambda_{12}\lambda_{21}}{\sqrt{(\lambda_{11}-\lambda_{22})^2+4\lambda_{12}\lambda_{21}}} [\ln t + U(h)][\ln t + U(\eta h)]\nonumber\\
+\left[1+\frac{\lambda_{11}-\lambda_{22}}{\sqrt{(\lambda_{11}-\lambda_{22})^2+4\lambda_{12}\lambda_{21}}}\right][\ln t + U(h)]+\nonumber\\
\left[1-\frac{\lambda_{11}-\lambda_{22}}{\sqrt{(\lambda_{11}-\lambda_{22})^2+4\lambda_{12}\lambda_{21}}}\right][\ln t + U(\eta h)]=0,
\end{eqnarray}
where $t=T/T_c$, $h=H_{c2}(\theta,T)D_1(\theta)/2\phi_{0}T$, $\eta=D_2(\theta)/D_1(\theta)$, $U(x)=\psi(1/2+x)-\psi(1/2)$, $\psi(x)$ is the di-gamma function, and $\phi_0$ is the flux quantum.
According to this model, in order for the contribution from a band to become negligible at low temperatures, its diffusivity must be much much less than that of the dominant band\cite{AGurevich2003}. Using this model, the strong curvature of $H_{c2}^{\perp}(T)$ can only be fit if the two bands have approximately equal diffusivity in this direction. Thus the negative second derivative of $H_{c2}^{\perp}(T)$ (Fig.~\ref{Hc2Tdep}) indicates that two bands with similar diffusivity in this direction should be considered. However, if the isotropic band had much higher diffusivity for $H \parallel c$, one would expect a strong upward curvature of $H_{c2}^{\parallel}(T)$ at low temperatures rather than the downward curvature that is observed.  Therefore, the only solution that can fit the temperature dependence in both directions self-consistently has two bands with opposite anisotropy. In this situation multiple solutions still remain. Although, the ratios of diffusivities of the bands for $H\parallel c$ and $H\perp c$ are set, the anisotropies of the individual bands cannot be determined without angular dependence measurements. The sharpness of the maximum of $H_{c2}(\theta)$ for $H \perp c$ depends on the anisotropies of the bands, the relative diffusivities, and on the difference between $\lambda_{11}$ and $\lambda_{22}$. Indeed, other FeAs compounds show angular dependences more sharply peaked than expected from single band anisotropy \cite{JJaroszynski2008b}. Including measurements of $H_{c2}$ as a function of angle (Fig.~\ref{angular}) allows us to find $H_{c2}(\theta,T)$ and thus uniquely determine the relationship between the diffusivity constants at all angles. The only way to fit both the temperature dependence and the 10~K angular dependence is for the more anisotropic band to have ``$H_{c2}^{\parallel}$''$>$``$H_{c2}^{\perp}$.'' The parameters obtained from the fit predict an angular dependence that has two maxima at low temperature, with the minimum $H_{c2}$ at intermediate angles.  Indeed, we observe such behavior at 1.4~K. $H_{c2}$ decreases as $H$ is inclined away from the c-axis until a minimum is reached around 40\textdegree\ and then increases again as $H$ becomes perpendicular to the c-axis (see Figs.~\ref{angular} and~\ref{angularfits}). Measurements at 0.5~K show that $H_{c2}(50\textrm{\textdegree})$ is less than $H_{c2}^{\parallel}$ or $H_{c2}^{\perp}$, consistent with the minimum at intermediate angle observed at 1.4~K. Both the dashed line and the solid line in Fig.~\ref{angularfits} come from parameters that fit the temperature dependence well, but only the solid line fits the angular dependence at 10~K well.

Including the 1.4~K data in the fit, we can refine the parameters and place additional restrictions on the relationships between lambda values.  From the fit we determine that $\lambda_{22}/\lambda_{11} \approx 0.42$, with $\lambda_{12}=\lambda_{21}=0.23\lambda_{11}$. The off diagonal elements (interband coupling constants) should be less than $\lambda_{22}$, but they do not have to be equal to each other as only their product appears in Eq.~2. However, we can rule out the possibility of the off diagonal matrix elements being $\ll \lambda_{22}$.  We obtain similar diffusivities for $H \perp c$ (i.e., $\sqrt{D_1^{(a)}D_1^{(c)}}=1.66\sqrt{D_2^{(a)}D_2^{(c)}}$) and a diffusivity ratio $D_1^{(a)}/D_2^{(a)} \approx 4.3$ for $H \parallel c$ with the more anisotropic band having lower diffusivity and hence higher ``$H_{c2}$'' when $H \parallel c$.  Thus the band anisotropies are 1.35 and 0.52 (i.e., ``$H_{c2}^{\parallel}$''$/$``$H_{c2}^{\perp}$'' would be 2, for that band), respectively. The best fit predicts sharper double maxima in $H_{c2}(\theta)$ at 1.4~K than are observed (Fig.\ref{angularfits}). The $\rho$ vs.\ $H$ curves are approximately parallel above 20\% of $\rho_n$ (Fig.~\ref{angular}), so there is little criteria-dependence. At low dissipation there is a clear difference which may be explained by enhanced vortex pinning perpendicular to the c-axis and/or the tail of an $H_{c2}$ distribution due to doping or strain, however that discussion is outside of the scope of this letter and will be addressed elsewhere.

These anisotropy parameters are surprising, as the band structure of undoped \ce{AEFe2As2} is composed of parallel, corrugated cylinders \cite{DJSingh2008,CLiu2008,VBZabolotnyy2008,SESebastian2008}. The more strongly corrugated hole band is expected to shrink with electron doping, and hence \ce{Sr(Fe , Co)2As2} may be more anisotropic than the hole-doped \ce{(Ba , K)Fe2As2} \cite{CLiu2008,LZhao2008}. However, there are a few scenarios compatible with both these band-structure models and our results that explain why the parameters are not as anticipated. Electron doping could cause the hole band to become ellipsoidal before vanishing, resulting in a band with a Fermi velocity oriented along the c-axis (consistent with Ref.~~\onlinecite{SESebastian2008}). Also, the complex bottle shape of the cylinders not only makes them less two dimensional, but it also means that the elliptical approximation used in this model may not capture important details of the Fermi surface.  Another possibility is that more than two bands contribute significantly to the properties of this material. Angle-resolved photoemission spectroscopy data, and band structure calculations indicate the existence of more than two bands \cite{HDing2008,DJSingh2008,CLiu2008,LZhao2008,HIkeda2008,KKuroki2008,KNakamura2008,VBZabolotnyy2008,SESebastian2008}, and this model only incorporates two bands. In addition the anisotropy of the bands may not be exactly as predicted because strain and doping differ from the parameters used in the band structure calculations.  The ratios between components of the lambda tensor obtained from the fit are within 20\% of that calculated for \ce{MgB2} \cite{AAGolubov2002}, except that in this case the more isotropic band has a larger (rather than smaller) intraband coupling than the more anisotropic band.  While the two band model fits our data fairly well, it is not possible to fully reproduce the sharp curvature of $H_{c2}^{\perp}(T)$ (even if the diffusivities of the two bands are taken to be equal). This results in a fit that only approximates the temperature dependence of $\gamma$, showing a flatter dependence and lower anisotropy near $T_c$. However, one should not expect a relatively simple model to perfectly fit the behavior of a superconductor with such a complex band structure.  Nevertheless, this model captures the essence of the behavior of $H_{c2}$.

While this model gives somewhat surprising results considering the expected band structure, it is still a reasonable approximation of the complex band structure of these materials. The similar importance of both bands points to a rich range of possible behavior available with small changes in structure or doping. Perhaps in these materials it is possible to obtain nearly isotropic behavior at low temperature while retaining the high $T_c$ and high $H_{c2}$ typically associated with more anisotropic compositions. This combination of low anisotropy, high $T_c$, and high $H_{c2}$ is highly desirable for applications, because it can result in high critical currents at any field orientation. It is important to note that both anisotropy parameters obtained from the fit are low in comparison to high temperature superconductors. The unusual anisotropy observed for the films in this study suggests the possibility of an \ce{AEFe2As2} superconductor with an $H_{c2}$ `anisotropy' ($\gamma$) that reverses at lower temperatures.

%\begin{acknowledgments}
Grant-in-Aid JSPS (Grant No.\ 19.9728) provided support for Y. Kohama. This work is also supported by an NHMFL UCGP grant, the U. S. NSF, the U.S. DOE, and the state of Florida.
%\end{acknowledgments}
%% Create the reference section using BibTeX:

\end{document}